\documentstyle[12pt]{article}
\newcommand{\beq}{\begin{equation}}
\newcommand{\eeq}{\end{equation}}
\newcommand{\bea}{\begin{eqnarray}}
\newcommand{\eea}{\end{eqnarray}}

\newcommand{\p}{\phi}

\renewcommand{\d}{\delta}

\renewcommand{\L}{\Lambda}

\newcommand{\n}{\nu}
\newcommand{\m}{\mu}

\newcommand{\th}{\theta}

\newcommand{\oh}{\frac{1}{2}}

\newcommand{\non}{\nonumber}
\newcommand{\zp}{\frac{d^Dp}{(2\pi)^D}}

\newcommand{\rf}[1]{(\ref{#1})}
\newcommand{\ra}{\rightarrow}
\begin{document}

\addtolength{\baselineskip}{0.20\baselineskip}
\hfill    NBI-HE-92-59

\hfill August 1992
\begin{center}

\vspace{36pt}
{\large \bf DYNAMICAL ORIGIN OF THE LORENTZIAN SIGNATURE OF SPACETIME}
\footnote{Supported by the U.S. Department of Energy under Grant
No. DE-FG03-92ER40711.}

\end{center}

\vspace{36pt}

\begin{center}
{\sl J. Greensite}
\footnote{Permanent address: Physics and Astronomy Dept.,
San Francisco State University, San Francisco, CA 94132 USA.
Email: greensit@stars.sfsu.edu} \\

\vspace{12pt}

The Niels Bohr Institute\\
Blegdamsvej 17\\
DK-2100 Copenhagen \O , Denmark\\

\end{center}

\vfill

\begin{center}
{\bf Abstract}
\end{center}

\vspace{12pt}

    It is suggested that not only the curvature, but also the
signature of spacetime is subject to quantum fluctuations.
A generalized D-dimensional spacetime metric of the form
$g_{\mu \nu}=e^a_\mu \eta_{ab} e^b_\nu$ is introduced, where
$\eta_{ab} = diag\{e^{ i\theta},1,...,1\}$.  The corresponding
functional integral for quantized fields then interpolates from a
Euclidean path integral in Euclidean space, at $\theta=0$, to a
Feynman path integral in Minkowski space, at $\theta=\pi$.  Treating
the phase $e^{i\theta}$ as just another quantized field, the signature
of spacetime is determined dynamically by its expectation value.
The complex-valued effective potential $V(\th)$ for the phase field,
induced by massless fields at one-loop, is considered.  It is argued that
$Re[V(\th)]$ is minimized and $Im[V(\th)]$ is stationary,
uniquely in $D=4$ dimensions, at
$\theta=\pi$, which suggests a dynamical origin for the Lorentzian
signature of spacetime.

\vspace{24pt}

\vfill

\newpage

 Spacetime curvature is a dynamical object in gravity theory;
spacetime signature is not.  With few exceptions \cite{Hartle,Tata},
the Lorentzian signature of the physical spacetime metric is simply
taken as given and non-dynamical.  Lorentzian signature can be enforced in
quantum gravity by introducing tetrads $e^a_\m$, so that

\beq
       g_{\m \n} = e^a_\m \eta_{ab} e^b_\n
\eeq
is the spacetime metric and

\beq
       \eta_{ab} = diag\{-1,1,1,1\}
\label{Mink}
\eeq
is the local frame Minkowski metric.  The problem in relativistic quantum
theory is then to evaluate Feynman path integrals of the form

\beq
       Z_F = \int d\m(e,\p,\psi,\overline{\psi}) \;
\exp\left[-i \int d^4x \sqrt{-g} \cal{L}
\right]
\label{ZF}
\eeq
with $\eta_{ab}$ fixed, where $d\m(e,\p,\psi,\overline{\psi})$
is the integration measure for
the tetrads, and other bosonic ($\p$) and fermionic ($\psi,\overline{\psi}$)
fields.

  For technical reasons, however, one often considers instead
the Euclidean path integral

\beq
       Z_E = \int d\m(e,\p,\psi,\overline{\psi}) \;
\exp \left[- \int d^4x \sqrt{g} \cal{L} \right]
\label{ZE}
\eeq
where this time

\beq
        \eta_{ab} = diag\{1,1,1,1\}
\label{Euclid}
\eeq
Comparing the Feynman and Euclidean path integrals, it is easy
to write down a slightly more general path integral which interpolates
between them, namely (in D dimensions)

\beq
      Z = \int d\m(e,\p,\psi,\overline{\psi}) \;
\exp\left[- \int d^Dx \sqrt{g} \cal{L}
\right]
\label{ZG}
\eeq
where
\beq
        \eta_{ab} = diag\{e^{i\th},1,...,1\}
\label{general}
\eeq
and we obtain the Euclidean theory for $\th=0$, and the
Feynman theory for $\th=\pi$, with
the correct $i \epsilon$ prescription for propagators
supplied automatically as $\th \ra \pi$.  The theory at $\th \ra - \pi$
converts to the Feynman theory under a time inversion.

  It should be stressed that for quantum gravity, the continuation from
the \linebreak
Minkowski to the Euclidean theory is really a continuation
in signature $\eta_{ab}$, rather than just a rotation $t \ra e^{i\th}t$
of the time coordinate.  Without continuation in the signature, the local
frame invariance of general relativity would be $O(3,1)$ in both Minkowski
{\it and} Euclidean space, instead of changing from $O(3,1)$ to $O(4)$
in Euclidean space.  Moreover, expectation values of certain
diffeomorphism invariant quantities (such as $<\int \sqrt{g} R^p>$)
have no dependence whatever on time intervals, and the
difference between expectation values in the Euclidean and
Minkowski theories resides entirely in their dependence on the
determinant of the signature $det(\eta)$.

  The introduction of a generalized signature \rf{general} then
suggests the possibility of viewing the phase
factor $\exp(i\th)$ as a dynamical field in its own right.  In that case,
the signature of spacetime will be determined dynamically, by the
expectation value of this phase field.  To compute the signature of
spacetime, we need to compute the effective potential $V(\th)$ for the
$\exp[i\th(x)]$ phase field, which is generated after integrating out all
other fields - matter, gauge, and tetrad.  Apart from the form of the
Langrangian, the final answer also calls for some assumptions about the
$\th$-dependence of the integration measure $d\m(e,\p,\psi,\overline{\psi})$,
which is otherwise taken proportional to the (real-valued) DeWitt measure.
In this letter I will just point out the consequences of the following
simple assumptions about the measure, which fix this $\th$-dependence:

\begin{description}
\item[~~1.] ~For free fields of mass $m$, the contributions to
$Z$ in eq. \rf{ZG} from
each (propagating) bosonic degree of freedom are equal, and inverse to
the contributions from each fermionic degree of freedom.  Thus, e.g.,
$Z=1$ at any $\th$ for a supersymmetric combination of free fields.
\item[~~2.] ~The integration measure for scalar fields is given by the
real-valued, invariant volume measure (DeWitt measure) in superspace
$d\m(\p) = D\p \sqrt{|G|}$, where $G$ is the determinant of the
scalar field supermetric $G(x,y)=\sqrt{g}\d(x-y)$.
\end{description}

  Consider the one-loop contribution to $V(\th)$ due to
integration over a massless scalar field $\p$ in a flat background
$e^a_\m=\d^a_\m$. Denoting this contribution by $V_0(\th)$, we
have

\beq
      \exp[-\int d^Dx V_0(\th)] = det^{-\oh}[- \sqrt{\eta}\eta^{ab}
\partial_a \partial_b]
\eeq
and proper-time regulation of the determinant gives

\bea
    V_0(\th) &=& -\oh \int^{\infty}_{1/\L^2} {ds \over s}\int \zp
\; \exp[- s(e^{-i\th/2}p_0^2 + e^{i\th/2}\vec{p}^{\;2})]
\non \\
       &=& -{\L^D \over D(4\pi)^{D/2}} \exp[-i(D-2)\th/4]
\label{proper}
\eea
where $\L$ is a momentum cutoff which, given the non-renormalizability
of gravity, presumably exists at the Planck scale.  The $p$-integration
in \rf{proper} is only well-defined for $\th$ in the range $[-\pi,\pi]$,
which is related to the fact that $Re[\sqrt{g} \cal{L_\p}]$ for a scalar
field is only bounded from below for $|\th| \le \pi$.\footnote{For the same
reason, the usual Wick rotation from Minkowski to Euclidean time must
be taken as $t \ra -i t$ rather than $t \ra +i t$.}

   For higher-spin massless fields, it is straightforward to verify
that, up to some extra factors of $det^p(\eta)$, the one-loop contribution
from each massless bosonic field is given by $det^{-\oh}(-\sqrt{\eta}
\eta^{ab}\partial_a \partial_b)$ raised to the power $n_B$ (the number
of propagating degrees of freedom),\footnote{E.g. $n_B=(D-2)$
degrees of freedom for massless vector fields; $n_B=D(D-3)/2$
for the graviton field.}
while for spinor fields it is this quantity raised to the power
$-n_F$ (no. of fermionic degrees of freedom $\times -1$).  In a curved
space-time background, the only difference is that the argument of the
root determinant changes to  $-\sqrt{g}g^{\m \n}\partial_\m \partial_\n$.
Any additional factors of $det^p(\eta)$ that arise in the integration are,
by assumption, cancelled by a corresponding factor in the measure.
Therefore, taking all massless fields into account, we have

\beq
    V(\th) =  (n_F-n_B){\L^D \over D(4\pi)^{D/2}} \exp[-i(D-2)\th/4]
\label{V}
\eeq
as the one-loop effective potential for the phase field.

   $V(\th)$ is complex-valued.  To determine $<e^{i\th}>$ at one-loop
level, we look for a value of $\th$ in the range $[-\pi,\pi]$
such that: i) $Re[V]$ is minimized; and ii) $Im[V]$ is stationary.
If these two conditions are not satisfied for the same value
of $\th$, then there may be large quantum fluctuations
in the signature.  From \rf{V}, the requirements are seen to
be:

\bea
    \left. \begin{array}{rr}
          \cos{[(D-2)\th/4]} = 0 \\
                                     \\
         \min{[Re[V(\th)]]}= 0     \\
  \end{array} \right\} \mbox{~~~$\th \in [-\pi,\pi]$}
\label{conditions}
\eea
The first condition is just the requirement that $Im[V]$ is
stationary.  Then $Re[V]=0$ at the stationary point of $Im[V]$, and
the second condition is the requirement that this is also the
minimum of $Re[V]$.  The constraint on the range of $\th$
is needed, as mentioned above, to ensure the existence of the
functional integral over the scalar field.

   The set of conditions \rf{conditions} cannot be solved in arbitrary
dimensions; in fact, for $n_F<n_B$ there is no solution in any
dimension.  This is because, for $n_F<n_B$, $\min[Re[V(\th)]]=V(0)\ne
0$.  For $n_F>n_B$ there is only one solution, namely,
$D=4$ dimensions and $\th=\pm \pi$.  This can be seen from the fact that
if $(D-2)\pi/4 > \pi/2$, then $\min[Re[V(\th)]]<0$, and similarly
if $(D-2) \pi/4 < \pi/2$, the minimum of the real part is greater than
zero.  Only at $(D-2)\pi/4 = \pi/2$, i.e. at $D=4$, can all conditions
be satisfied, and this is just at $\th=\pm \pi$. Remarkably,
{\it it appears that $V(\th)$ uniquely
singles out both Lorentzian signature and the observed
dimensionality of spacetime.}

   In this argument I have neglected the fact that the
gravitational action is unbounded from below for any $\theta$,
due to the "wrong-sign" of the kinetic term for the conformal factor.
The one-loop contribution from the gravitational field to $V(\th)$ is
therefore ill-defined.  There are a number of approaches to defining
the Euclidean path integral for gravity, such as stochastic stabilization
\cite{Me}, contour rotation \cite{GHP}, and other ideas
\cite{Others}.  But whichever prescription is used, so long as
the eigenvalues of the graviton kinetic operator are
proportional to those of the Klein-Gordon operator $\eta^{ab}
\partial_a \partial_b$, the above conclusion concerning signature and
dimension is unaffected.

   The possibility that spacetime signature might fluctuate raises
many questions.  Conceivably such fluctuations would be important
at the Planck scale, or relevant to the last stages of gravitational
collapse.  Supersymmetry is also a concern since, at the one-loop level,
Lorentzian signature arises at D=4 only if $n_F>n_B$.\footnote{In
connection with the role of massless spinors, I have learned that
Nielsen \cite{Holger} has recently arrived at the same conclusion
regarding signature and dimension based on properties of chiral fermions,
although his reasoning is quite different from the argument presented here.}
If Nature is supersymmetric ($n_F=n_B$), then the expectation value of the
signature presumably depends on the details of the supersymmetry breaking.
This issue requires further study.

   The Euclidean $\rightarrow$ Minkowski interpolation takes us from a
real measure, $\exp(-S)$ to a complex measure, i.e. from statistical
mechanics to quantum mechanics.  An obvious
further generalization would be to have $\eta_{ab}$ interpolate between
all possible signatures, with arbitrary numbers of $\pm$ signs corresponding
to arbitrary numbers of time-like coordinates.  However,
quantum theory with more than one time variable has not,
to my knowledge, been formulated, and it is not obvious what the
correct measure for the corresponding functional integral should be.
One could speculate that interpolation to quantum mechanics with
multiple time-like coordinates might call for some generalization of
complex numbers, such as the quaternions or octonions; but at present
it is unclear if multi-time generalizations of quantum theory can be
formulated consistently.

\bigskip

   To summarize, if the signature of the metric is a dynamical quantity,
then its expectation value is determined by quantum effects. In this
letter I have discussed the one-loop effective potential for the
generalized signature \rf{general} induced by
massless fields, given certain assumptions about the
functional integration measure.  It is found that if the number
of fermionic degrees of freedom exceeds the number of bosonic degrees
of freedom, then the
real part of the effective potential is minimized (and the imaginary
part is stationary) only for Lorentzian signature, and only in D=4
dimensions.

\vspace{33pt}

\noindent {\Large \bf Acknowledgements}{\vspace{11pt}}

    I thank Marty Halpern, John Moffat, Holger Bech Nielsen, and Niels
Obers for helpful discussions.

\end{document}